\RequirePackage{fix-cm}
\documentclass[twocolumn,epjc3]{svjour3}  
\usepackage{enumitem}
\usepackage{graphicx}
\usepackage{subfigure}
\usepackage{dcolumn}
\usepackage{array,multirow}
\usepackage{bm}
\usepackage{amsmath}
\usepackage{epstopdf}
\usepackage{amsfonts}
\usepackage{amssymb}
\usepackage{epsfig}
\usepackage{tabularx}
\usepackage{color}
\usepackage[colorlinks=blue,citecolor=blue,urlcolor=blue,breaklinks]{hyperref}
\smartqed  
\RequirePackage{graphicx}
\journalname{Eur. Phys. J. C}
\begin{document}

\title{Baryogenesis via leptogenesis in multi-field inflation}

\author{
Grigoris Panotopoulos\thanksref{e1,addr1}
\and
Nelson Videla\thanksref{e2,addr2} 
}


\thankstext{e1}{grigorios.panotopoulos@tecnico.ulisboa.pt}
\thankstext{e2}{nelson.videla@pucv.cl}

\institute{
Centro de Astrof\'{\i}sica e Gravita\c c\~ao-CENTRA, Departamento de F\'{\i}sica, Instituto Superior T{\'e}cnico - IST, Universidade de Lisboa-UL, Av. Rovisco Pais 1, 1049-001 Lisboa, Portugal\label{addr1}
\and
Instituto de F\'{\i}sica, Pontificia Universidad Cat\'{o}lica de Valpara\'{\i}so,
Avenida Brasil 2950, Casilla 4059, Valpara\'{\i}so, Chile\label{addr2}
}

\date{Received: date / Accepted: date}

\maketitle

\begin{abstract}
In multi-field reheating after modular $j$-inflation we investigate the conditions under which baryogenesis via non-thermal leptogenesis can be successfully realized. We introduce three heavy right-handed neutrinos to the non-supersymmetric Standard Model of particle physics, assuming hierarchical neutrino masses. Considering a typical mass for the first right-handed neutrino of the order of $10^{11}~GeV$, suggested from the seesaw mechanism and also from concrete $SO(10)$ grand unification models, we obtain the allowed parameter space for viable baryogenesis. An upper bound for the inflaton mass as well as a lower bound for its branching ratio to the pair of lightest right-handed neutrinos are found and reported.
\end{abstract}

\maketitle

\section{Introduction}

Inflation \cite{starobinsky1,inflation} is widely accepted as the standard paradigm of the early Universe. The first reason is due to the fact that several long-standing puzzles of the Hot Big-Bang model, such as the horizon, flatness, and monopole problems, find a natural explanation in the framework of inflationary Universe. In addition, and perhaps the most intriguing feature of inflation, is that it gives us a causal interpretation of the origin of the Cosmic Microwave Background (CMB) temperature anisotropies \cite{Abazajian:2013vfg}, while at the same time it provides us with a mechanism to explain the Large-Scale Structure (LSS) of the Universe, since quantum fluctuations during the inflationary era may give rise to the primordial density perturbations \cite{mukhanov}.

Although single-field slow-roll inflation provides us with the best fit to the data, considering multi-field inflation offers a wide range of
new features, which go beyond the predictions of single-field scenarios, and which could be detected. As a few examples, we may mention isocurvature perturbations, correlated or anti-correlated with the curvature perturbation, and detectable level of non-Gaussianity. In particular, for the current observational constraints on isocurvature perturbations, see Ref.\cite{Akrami:2018odb}. More fundamentally, theories where the interactions of the Standard Model particles are unified with gravity, such as supergravity \cite{nilles} and Superstring theory \cite{ST1,ST2}, give rise to multiple fields instead of one. For a comprehensive review on multi-field inflation, see e.g.\cite{Gong:2016qmq,Langlois:2010xc}.

The Standard Model (SM) of particle physics has been extremely successful describing very accurately a vast amount of observational data at energies that span many orders of magnitude. Despite its success, however, it is widely accepted that the SM is the low energy limit of some underlying fundamental theory. Perhaps the most straightforward evidence for physics beyond the SM is the tiny neutrino masses in the sub-eV range seen in solar, atmospheric, reactor and accelerator neutrino experiments \cite{SK1,SK2,SK3,SNO,kamland,chooz,k2k}. Right-handed neutrinos are very well motivated hypothetical particles postulated to exist due to their appealing properties, which can be summarized as follows: they can explain small neutrino masses via the seesaw mechanism \cite{seesaw}, they fit very nicely in the spinorial 16-dimensional representation of $SO(10)$ GUT group \cite{unification}, and finally they can explain the baryon asymmetry in the Universe via leptogenesis, see the next paragraph.

One of the goals of successful inflation must be the explanation of the baryon asymmetry in the Universe, which comprises on of the biggest challenges of modern cosmology. Both primordial Big Bang nucleosynthesis \cite{bbn} and data from the CMB temperature anisotropies \cite{wmap,planck1,planck2} show that the baryon-to-photon ratio is a tiny number, $\eta_B=6.19 \times 10^{-10}$ \cite{values}. This number should be calculable within the framework of known particle physics. Although several mechanisms exist, perhaps the most elegant one is leptogenesis \cite{leptogenesis}. A lepton asymmetry via the out-of-equilibrium decays of right-handed neutrinos is generated first, and then this lepton asymmetry is partially converted into baryon asymmetry via non-perturbative "sphaleron" effects \cite{sphalerons}. 

Of particular interest is the non-thermal leptogenesis scenario \cite{lepto1,lepto3,lepto4,lepto5,lepto6,lepto7}, since the lepton asymmetry within the framework of non-thermal leptogenesis is proportional to the reheating temperature after inflation. The two key parameters of the big bang, namely the baryon asymmetry and the reheating temperature, are therefore linked together. Recently in \cite{multifield} the author studied multifield reheating in interacting theories where the inflaton trajectory is weakly curved, and he realized this scenario in a particular example of modular inflation. It is the goal of this article to investigate the conditions under which baryogenesis via non-thermal leptogenesis in the scenario of multifield reheating considered in \cite{multifield} can be successfully realized. Our work is organized as follows: In the next section we briefly present the theoretical framework, while the numerical results are discussed in section three. Finally we conclude summarizing our main findings in the last section.

\section{Theoretical framework}

\subsection{Modular $j$-inflation}

We briefly present modular $j$-inflation. For more details the interested reader may consult \cite{multifield,modular1,modular2}. In the multifield scenario there are a bunch of inflaton-type fields $\phi^I$, and another bunch of decay products $\chi^A$. The total Lagrangian density takes the usual form
\begin{equation}
\mathcal{L} = \mathcal{L}_{kin} - \mathcal{L}_{int}
\end{equation}
where the non-trivial kinetic term $\mathcal{L}_{kin}$ is given by
\begin{equation}
\mathcal{L}_{kin} = \frac{1}{2} G_{IJ} g^{\mu \nu} \partial_\mu \phi^I \partial_\nu \phi^J + \frac{1}{2} G_{AB} g^{\mu \nu} \partial_\mu \chi^A \partial_\nu \chi^B
\end{equation}
with $G_{IJ},G_{AB}$ being the metric tensors in the field space for the $\phi$ fields and the $\chi$ fields, respectively, while the interaction term is given by
\begin{equation}
\mathcal{L}_{int} = V(\phi^I) + U(\chi^A) + W(\phi^I,\chi^A)
\end{equation}
with a potential $V$ for the $\phi$ fields only, another potential $U$ for the $\chi$ fields only, and
an interaction term between the two types of scalar fields
\begin{equation}
W(\phi^I,\chi^A) = \frac{1}{2} \sum_{I,A} g_{I,A} \phi^I (\chi^A)^2 + \frac{1}{4} \sum_{I,A} h_{I,A} (\phi^I)^2 (\chi^A)^2
\end{equation}
with a coupling constant $g_{I,A}$ for the triscalar interactions (with dimensions of mass), and another dimensionless coupling constant $h_{I,A}$ for the bi-quadratic interactions, leading to decay and scattering processes of the form
\begin{equation}
\phi^I \rightarrow \chi^A \chi^A 
\end{equation}

\begin{equation}
\phi^I \phi^I  \rightarrow  \chi^A \chi^A
\end{equation}

Modular inflation, not to be confused with moduli inflation, where inflation is driven by moduli fields coming from Superstring compactifications on Calabi-Yau manifolds, is a two-field inflationary model with modular functions for the field target space. In modular $j$-inflation, a particular example of modular inflation, the field space has a non-trivial metric geometry characterized by the Poincar{\'e} metric
\begin{equation}
ds^2 = \frac{(d \tau^1)^2+(d \tau^2)^2}{(\tau^2)^2}
\end{equation}
with the two inflaton fields being $\phi^1=\mu \tau^1$ and $\phi^2=\mu \tau^2$, where $\mu$ is some mass scale. The inflaton potential is given by
\begin{equation}
V(\phi^I) = \Lambda^4 F(\phi^I/\mu)
\end{equation}
where $\Lambda$ is another mass scale, and $F(x)$ is a dimensionless function. The potential can be Taylor expanded around its minimum $\phi_0$, and assuming that the difference $\phi-\phi_0$ is small, the
inflaton potential may be approximated by a monomial of the form
\begin{equation}
V(\phi) \simeq \Lambda^4 \left( \frac{\phi-\phi_0}{\mu} \right)^2
\end{equation}
with $\phi=\phi^1$ being the dominant component along the inflationary trajectory, which is assumed to be only weakly curved. Therefore, during reheating the second inflaton field will be irrelevant, and the inflaton potential looks like the one of a usual single-inflationary model. Despite the similarity, however, the phenomenology of the scenario discussed here is different compared to genuine single-field inflationaty models, such as Natural Inflation \cite{NI1,NI2,NI3} or chaotic inflation based on a mass term for the inflaton, see a couple of comments in the end of the next section.

In the multi-field reheating realized in the modular scenario of \cite{multifield} the inflaton potential is computed to be
\begin{equation}
V(\phi) = \frac{\Lambda^4}{\mu^2} (\phi-\mu/2)^2
\end{equation}
leading to an inflaton mass
\begin{equation}
m_\phi = \frac{\sqrt{2} \Lambda^2}{\mu}
\end{equation}
In addition, the inflaton decays into bosonic decay products
$\chi$ via a triscalar interaction term in the Lagrangian density with a coupling constant $g$
\begin{equation}
\mathcal{L}_{\chi \chi \phi} = g \phi \chi^2
\end{equation}

\subsection{Non-thermal leptogenesis}

In the scenario of non-thermal leptogenesis, the lepton asymmetry is given by
\begin{equation}
Y_L = \frac{3}{2} BR(\phi \rightarrow N_1 N_1) \frac{T_{reh}}{m_\phi} \epsilon
\end{equation}
where $m_\phi$ is the mass of the inflaton, $\epsilon$ is the CP-violation asymmetry factor, $T_{reh}$ is the reheating temperature after inflation, and
$BR(\phi \rightarrow N_1 N_1) \equiv BR$ is the branching ratio of the inflaton decay channel into a pair of right-handed neutrinos $\phi \rightarrow N_1 N_1$.

The CP-violation asymmetry factor is defined by \cite{covi}
\begin{equation}
\epsilon = \frac{\Gamma-\bar{\Gamma}}{\Gamma+\bar{\Gamma}}
\end{equation}
where $\Gamma=\Gamma(N \rightarrow l H)$ and $\bar{\Gamma}=\Gamma(N \rightarrow \bar{l} H^\dagger)$
and it can be written down as
\begin{equation}
\epsilon = \epsilon_{max} sin \delta
\end{equation}
where the maximum CP-asymmetry factor (assuming hierarchical neutrino masses, $M_1 \ll M_2, M_3$) has been computed to be \cite{ibarra}
\begin{equation}
\epsilon_{max} = \frac{3}{8 \pi} \; \frac{M_1 \sqrt{\Delta m_{atm}^2}}{v^2}
\end{equation}
with $v=246~GeV$ being the Higgs vacuum expectation value, $M_1$ being the mass of the right-handed neutrino of the first family $N_1$, and $\Delta m_{atm}^2 = 2.5 \times 10^{-3}~eV^2$ being the atmospheric neutrino mass difference from neutrino oscillation data \cite{values}.

During reheating \cite{reh1,reh2,reh3} the particle production is taken into account via a phenomenological approach in which an additional term is added into the Klein-Gordon equation
\begin{equation}
\ddot{\phi} + (3 H + \Gamma_\phi) \dot{\phi} + \frac{dV}{d \phi}= 0
\end{equation}
where $\Gamma_\phi$ is the total inflaton decay. The reheating temperature after inflation is given by \cite{multifield}
\begin{equation}
T_{reh} = \left( \frac{90}{g_*} \right)^{1/4} \sqrt{M_{pl} \Gamma_\phi} 
\end{equation}
where $M_{pl}=2.4 \times 10^{18}~GeV$ is the reduced Planck mass, and $g_*$ counts for the relativistic degrees of freedom. In the SM $g_*=106.75$ \cite{covi}. The inflaton decays into right-handed neutrinos, and in the scenario of modular inflation realized in \cite{multifield} into bosonic decay products $\chi$ as well. Therefore the total inflaton decay width $\Gamma_\phi=\Gamma_f+\Gamma_b$ has two contributions, a fermionic one
\begin{equation}
\Gamma_f = \frac{|y_i|^2 m_\phi}{4 \pi}
\end{equation}
where $y_i$ is the Yukawa coupling of the inflaton to the right-handed neutrino $N_i$, and a bosonic one \cite{multifield}
\begin{equation}
\Gamma_b = \frac{g^2}{8 \pi m_\phi}
\end{equation}

By definition
\begin{equation}
BR = \frac{\Gamma_f}{\Gamma_{\phi}} = \frac{2 y_1^2 m_\phi^2}{2 y_1^2 m_\phi^2 + g^2}
\end{equation}
and we may easily solve for the Yukawa coupling to express $y_1$ in terms of $g, m_\phi$ for a given branching ratio as follows
\begin{equation}
y_1 = \frac{g}{\sqrt{2} m_\phi} \: \sqrt{\frac{BR}{1-BR}}
\end{equation}

Finally, the initial lepton asymmetry $Y_L=n_L/s$ is converted into baryon asymmetry $Y_B=n_B/s$ via sphaleron effects \cite{sphalerons}
\begin{equation}
Y_B = a Y_L 
\end{equation}
where $n$ is the number density of leptons or baryons, $s$ is the entropy density of radiation, $s=(2 \pi^2 h_* T^3)/45$, and the conversion factor $a$ is computed to be $a=(24+4N_H)/(66+13N_H)$ \cite{turner}, with $N_H$ being the number of Higgs doublets in the model. In the SM without supersymmetry with only one Higgs doublet $N_H=1$ and $a=28/79$.

We thus obtain the final expression for the baryon asymmetry
\begin{equation}
Y_B = a \: BR \; \frac{9}{16 \pi} \; \frac{T_{reh} M_1 \sqrt{\Delta m_{atm}^2}}{m_\phi v^2}
\end{equation}
assuming that $m_\phi > 2 M_1$, $sin \delta = 1$ (maximum $CP$ asymmetry factor), and that the other two channels $\phi \rightarrow N_2$ and $\phi \rightarrow N_3$ are kinematically closed. The baryon asymmetry is related to the baryon-to-photon ratio, and therefore $Y_B$ takes the observational value
\begin{equation}
Y_B = \frac{\eta_B}{7.04}=7.89 \times 10^{-11} \equiv Y_{obs}
\end{equation}
{Using the expressions for the total inflaton decay rate as well as the reheating temperature after inflation we may express $g$ as a function of the inflaton mass for a given branching ratio
\begin{equation}
g(m_{\phi}) = \frac{32}{9 \sqrt{15 M_p}} \: \left( \frac{\sqrt{10 g_*} \pi^3 m_{\phi}^3 v^4 Y_{obs}^2}{(a M_1 BR)^2 \Delta m_{atm}^2} \: (1-BR) \right)^{1/2}
\end{equation}
and accordingly for $y_1$
\begin{equation}
y_1(m_{\phi}) = \frac{g(m_{\phi})}{\sqrt{2} m_\phi} \: \sqrt{\frac{BR}{1-BR}}
\end{equation}

We see that $g$ and $y_1$ scale differently with the inflaton mass, namely $g \sim m_{\phi}^{3/2}$, while $y_1 \sim m_{\phi}^{1/2}$.

\section{Numerical results}

In the following we shall assume a typical value for the mass of the first right-handed neutrino
$M_1 = 10^{11}~GeV$ \cite{Fukuyama:2002ch}. Then the model is characterized by three free parameters, namely the Yukawa coupling $y_1$, the triscalar coupling constant $g$ and the inflaton mass $m_\phi$. It is more convenient, however, to work with the branching ratio $BR(\phi \rightarrow N_1 N_1)$ since it is a dimensionless number taking values in the range [0,1]. 

Imposing the observational constraint $Y_B=7.89 \times 10^{-11}$ we obtain the coupling constant $g$ as a function of the inflaton mass for different values of the branching ratio shown in Fig.~1. Note that in the usual baryogenesis via non-thermal leptogenesis scenario, at least in non-supersymmetric models, the inflaton decays into right-handed neutrinos only, there are no scalar decay products, and the branching ratio $BR \simeq 1$ (see e.g.\cite{lepto7}). Therefore we have considered here low branching rations up to 0.1. We see that $m_\phi \gg g$ as anticipated in \cite{multifield}. For each point of the curves shown in Fig.~1 we can compute the Yukawa coupling and the reheating temperature after inflation using the formulas presented in the previous section. 
We find that both $y_1$ and $T_{reh}$ increase with the inflaton mass, as shown in Fig.~2 and 3. For a given inflaton mass the reheating temperature after inflation decreases with the branching ratio. Therefore, the scenario studied here predicts a higher $T_{reh}$ compared to the standard discussion where $BR \simeq 1$.

Since the model is non-supersymmetric, there are no bounds on $T_{reh}$ due to the gravitino problem \cite{linde1,linde2}. Non-thermal leptogenesis, however, works if $T_{reh} \ll M_1$, and therefore we impose the condition $T_{reh} \leq M_1/100$, which implies an upper bound for the inflaton mass
\begin{equation}
m_\phi \leq \frac{9 a \sqrt{\Delta m_{atm}^2} M_1^2 BR}{1600 \pi v^2 Y_{obs}}
\end{equation}
proportional to the $BR$, and therefore for a viable baryobenesis via non-thermal leptogenesis the inflaton mass must take values in the range
\begin{equation}
2 M_1 < m_\phi \leq \frac{9 a \sqrt{\Delta m_{atm}^2} M_1^2 BR}{1600 \pi v^2 Y_{obs}}
\end{equation}
and this in turn implies a lower bound on the branching ratio, $BR \geq 0.003$. In figures 1-3 the upper bound of the inflaton mass is shown.

We see that the specific scenario with a low branching ratio discussed in this work requires a relatively light inflaton, $m_\phi < 10^{13}~GeV$, while in the chaotic inflationary model based on a mass term for the inflaton, $(1/2) m^2 \phi^2$, it is well-known that the COBE normalization requires an inflaton mass $m > 10^{13}~GeV$. Therefore, this single-field model cannot work here.

Finally, we may now show the mass scale $\mu$ as a function of the mass scale $\Lambda$ for a given value of the inflaton mass. But before that, since any viable inflationary model first should be compatible with the spectral index $n_{RR}$ and tensor-to-scalar ratio $r$ bounds, we briefly summarize here the main results obtained in \cite{pheno}. According to that work, within the framework of modular $j$-inflation the spectral index was found to be $n_{RR} = 0.96$, while $r$ was found to take values in the range $10^{-8} \leq r \leq 0.08$, compatible with the values reported by the Planck collaboration \cite{planck1,planck2} as well as the BICEP2/Keck/Planck collaboration \cite{bicep}. In addition, the scale $\Lambda$ is allowed to take values in the range $10^{-6} \leq \Lambda/M_p \leq 10^{-4}$, and it is lower than the mass scale in Natural Inflation, where it is of the order of the GUT scale \cite{NI1,NI2,NI3}.

In Fig.~4 we show the mass scale $\mu$ as a function of the mass scale $\Lambda$ for three different values of $m_\phi$. The scale $\Lambda$ takes values in its allowed range mentioned before, while the scale $\mu$ is of the order of the GUT scale.

\begin{figure}
\centering
\includegraphics[scale=0.35]{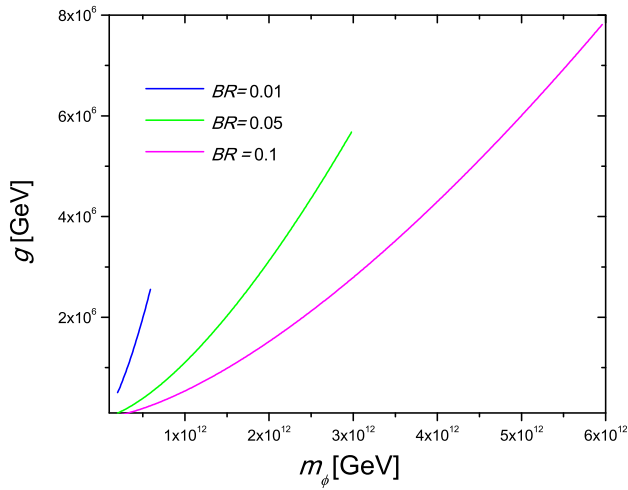}
\caption{Triscalar coupling constant $g$ versus the inflaton mass $m_\phi$ (both in GeV) for three different values of the branching ratio, 0.01 (blue), 0.05 (green) and 0.1 (magenta).}
\label{fig1}
\end{figure}

\begin{figure}
\centering
\includegraphics[scale=0.35]{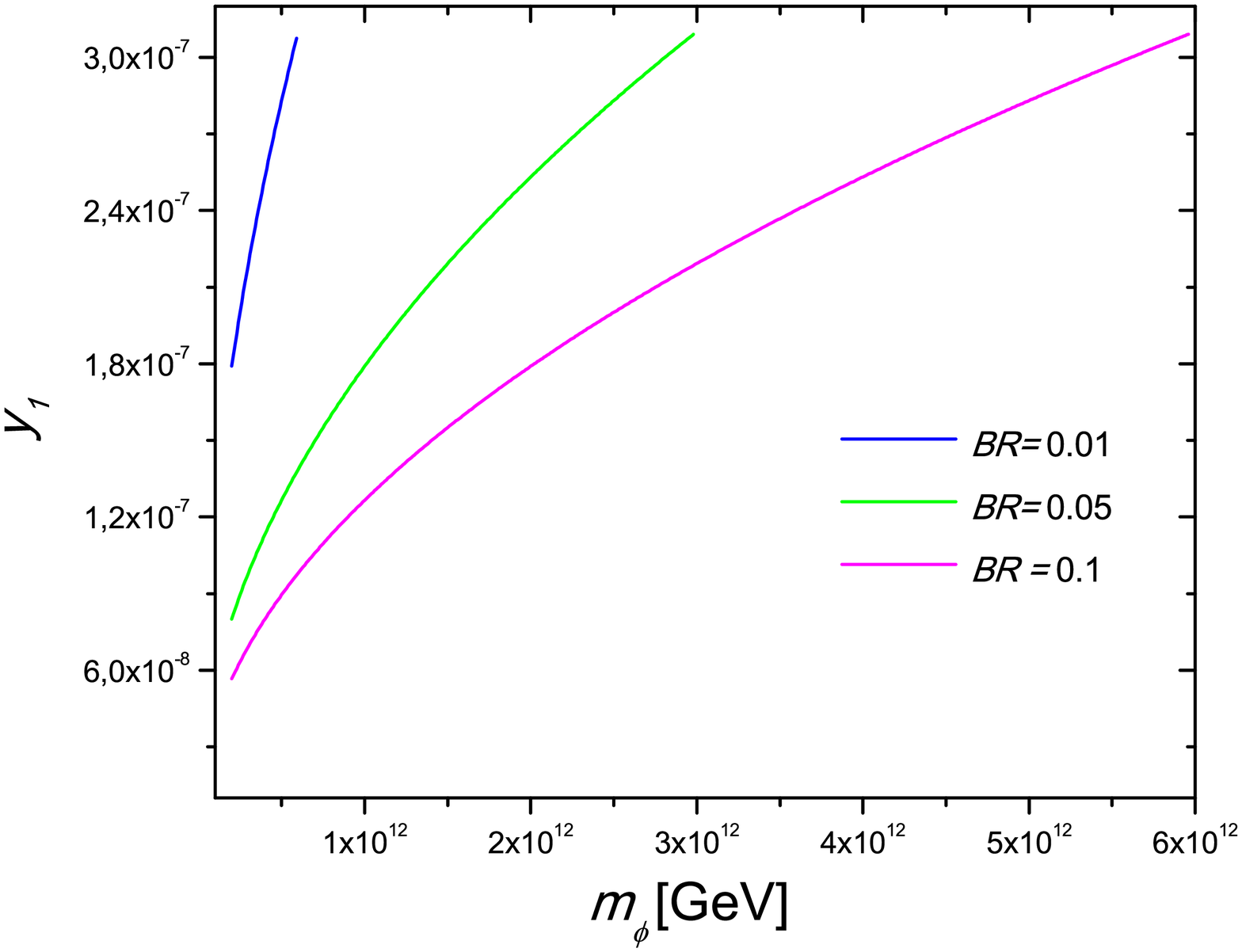}
\caption{Yukawa coupling $y_1$ versus the inflaton mass $m_\phi$ in GeV for three different values of the branching ratio, 0.01 (blue), 0.05 (green) and 0.1 (magenta).}
\label{fig2}
\end{figure}

\begin{figure}
\centering
\includegraphics[scale=0.35]{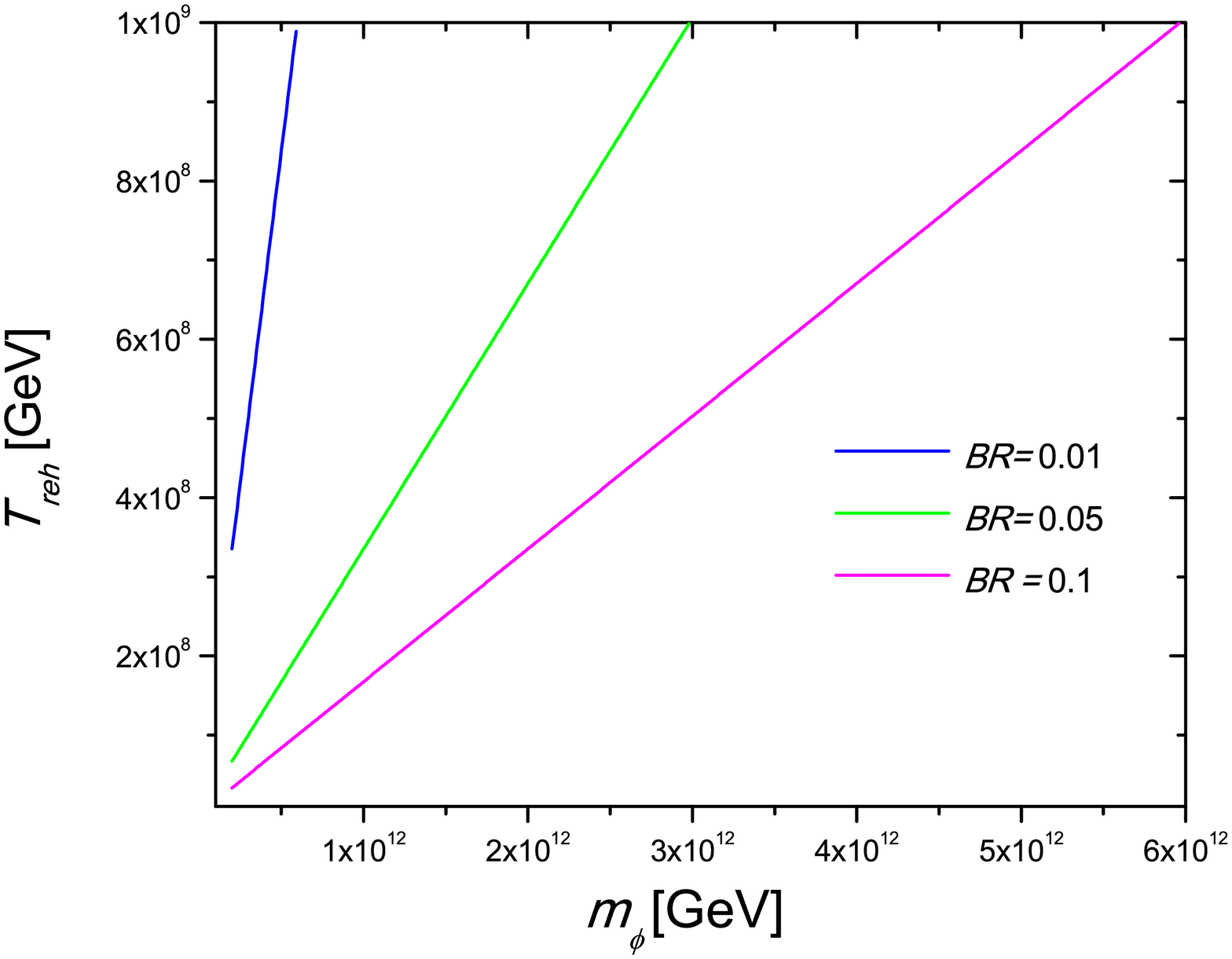}
\caption{Reheating temperature after inflation versus the inflaton mass mass (both in GeV) for three different values of the branching ratio, 0.01 (blue), 0.05 (green) and 0.1 (magenta).}
\label{fig3}
\end{figure}

\begin{figure}
\centering
\includegraphics[scale=0.45]{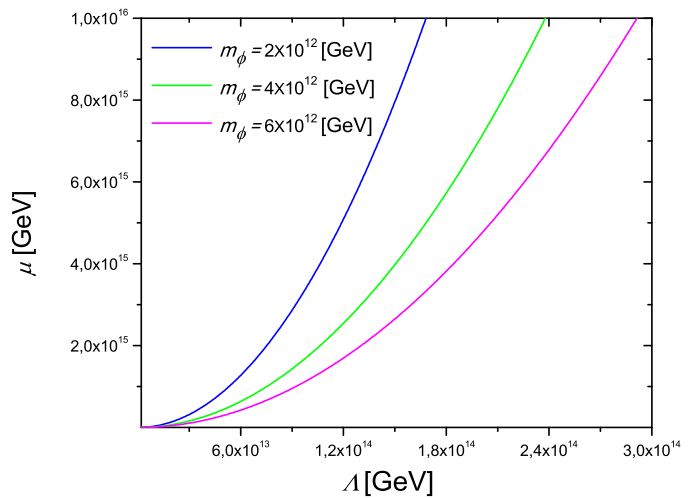}
{\vspace{-2 cm}
\caption{Mass scale $\mu$ versus mass scale $\Lambda$ (both in GeV) for three different values of the inflaton mass, $2 \times 10^{12}~GeV$ (blue), $4 \times 10^{12}~GeV$ (green) and $6 \times 10^{12}~GeV$ (magenta).}}
\label{fig4}
\end{figure}

\section{Conclusions}

To summarize, in this article we have studied baryogenesis via non-thermal leptogenesis in multi-field reheating realized in a particular example of modular inflation. We have assumed hierarchical neutrino masses in the seesaw mechanism scenario introducing three heavy right-handed neutrinos $N_i$ without supersymmetry, and we have investigated under what conditions the model is viable. The inflaton $\phi$ decays into heavy right-handed neutrinos, and into bosonic decay products too. We have focused to the case of a small branching ratio $\phi \rightarrow N_1 N_1$, and we have obtained the allowed parameter space corresponding to successful baryogenesis. An upper bound for the inflaton mass as well as a lower bound for its branching ratio into the pair of lightest right-handed neutrinos are obtained and reported. As a final remark,  we have assumed that perturbative reheating applies. Non-perturbative preheating effects after inflation deserve a more detailed analysis. We hope to be able to address this point in a future work.


\section*{Acknowlegements}

The author N.~V. was supported by Comisi\'on Nacional de Ciencias y Tecnolog\'ia of Chile through FONDECYT Grant N$^{\textup{o}}$ 11170162.
The author G.~P. thanks the Funda\c c\~ao para a Ci\^encia e Tecnologia (FCT), Portugal, for the financial support to the Center for Astrophysics and Gravitation-CENTRA, Instituto Superior T{\'e}cnico,  Universidade de Lisboa, through the Grant No. UID/FIS/00099/2013. He also wishes to thank the Pontificia Universidad Cat\'{o}lica de Valpara\'{\i}so, where part of the work was completed, for its warmest hospitality.


\end{document}